\documentclass[prb,superscriptaddress,showpacs,a4paper,twocolumn,floatfix]{revtex4}
\usepackage{amsmath} 
\usepackage{graphics} 
\usepackage{dcolumn}
\usepackage{epsfig} 
\usepackage{color} 
\usepackage{subfigure}
\usepackage[percent]{overpic}

\graphicspath{{figures/}}

\begin{document}

\title{Exploring the bonding of large hydrocarbons on noble metals:
  \\Diindoperylene on Cu(111), Ag(111), and Au(111)}

\author{C. B\"urker} 

\affiliation{Institut f\"ur Angewandte Physik, Universit\"at T\"ubingen, Auf der Morgenstelle 10, 72076
  T\"ubingen, Germany}
\author{N.~Ferri} \author{A.~Tkatchenko} 

\affiliation{Fritz-Haber-Institut der Max-Planck-Gesellschaft, Faradayweg 4-6,
  14195 Berlin, Germany}

\author{A.~Gerlach} 

\affiliation{Institut f\"ur Angewandte Physik, Universit\"at T\"ubingen, Auf
  der Morgenstelle 10, 72076 T\"ubingen, Germany}

\author{J.~Niederhausen} 

\affiliation{Institut f\"ur Physik, Humboldt-Universit\"at zu Berlin,
  Newtonstr. 15, 12489 Berlin, Germany}

\author{T.~Hosokai} 

\affiliation{Institut f\"ur Angewandte Physik, Universit\"at T\"ubingen, Auf
  der Morgenstelle 10, 72076 T\"ubingen, Germany}
\affiliation{Department of Materials Science and Engineering, Iwate University, Ueda
  4-3-5, Morioka 020-8551, Japan}

\author{S.~Duhm} 

\affiliation{Graduate School of Advanced Integration Science, Chiba
  University, Chiba 263-8522, Japan} 
\affiliation{Institute of Functional Nano
  \& Soft Materials (FUNSOM), Soochow University, 199 Ren-Ai Road, Suzhou 215123,
  P. R. China}

\author{J.~Zegenhagen} 

\affiliation{European Synchrotron Radiation Facility, 6 Rue Jules Horowitz, BP
  220, 38043 Grenoble Cedex 9, France}

\author{N.~Koch} 

\affiliation{Institut f\"ur Physik, Humboldt-Universit\"at zu Berlin,
  Newtonstr. 15, 12489 Berlin, Germany}

\author{F.~Schreiber} \email[Corresponding
author;\ ]{frank.schreiber@uni-tuebingen.de} 

\affiliation{Institut f\"ur Angewandte Physik, Universit\"at T\"ubingen, Auf
  der Morgenstelle 10, 72076 T\"ubingen, Germany}

\date{\today}

\begin{abstract}
  We present a benchmark study for the adsorption of a large $\pi$-conjugated
  organic molecule on different noble metal surfaces, which is based on x-ray
  standing wave (XSW) measurements and density functional theory calculations
  with van der Waals (vdW) interactions. The bonding distances of
  diindenoperylene on Cu(111), Ag(111), and Au(111) surfaces (2.51,
  3.01, and 3.10 \AA{}, respectively) determined with the normal
  incidence XSW technique are compared with calculations.  Excellent agreement
  with the experimental data, i.e.,\ deviations less than 0.1 \AA{}, is
  achieved using the Perdew-Burke-Ernzerhof (PBE) functional with vdW interactions
  that include the collective response of substrate electrons
  (the PBE+vdW$^{\rm{surf}}$ method). It is noteworthy that the calculations show that the
  vdW contribution to the adsorption energy increases in the order
  Au(111)~$<$~Ag(111)~$<$~Cu(111).
\end{abstract}

\maketitle

\section{Introduction}
\label{sec:intro}
%
%
The reliable prediction of the equilibrium structure and energetics of hybrid
inorganic/organic systems from first principles represents a great challenge
for theoretical methods due to the interplay of covalent interactions,
electron transfer processes, Pauli repulsion, and van der Waals (vdW)
interactions. During recent years, huge efforts have been made to incorporate
vdW interactions into density functional theory (DFT) calculations in order to
determine the structure and stability of $\pi$-conjugated organic molecules on
solid
surfaces~\cite{atodiresei_prl09,mercurio_prl10,stradi_prl11,olsen_prl11,McNellisPhD2010,tkatchenko_mrs10}.
Understanding these interface properties is relevant, \textit{inter alia}, for
electron transfer processes in organic devices.
%
%
Until now and despite the obvious benefit, there are only few a studies of
metal-organic interfaces combining theory and experiment. Here, x-ray standing wave
(XSW) measurements can provide an important test for DFT
calculations~\cite{ruiz_prl12,mercurio_prl10}. This is particularly important
for systems with strong vdW contributions to the overall bonding, for which no
simple substrate dependence is expected.
%
%
%
%

As model system we chose diindenoperylene (DIP,
$\mathrm{C}_{32}\mathrm{H}_{16}$), a $\pi$-conjugated organic semiconductor
with excellent optoelectronic device performance, which has been studied over
the last decade both in
thin-films~\cite{duerr_prl03,kowarik_prl06,heinemeyer_prl10,wagner_afm10} and
in monolayers on noble metal
surfaces~\cite{Oteyza_pccp09,Huang_pccp11,Oteyza_jpcc08}.
With respect to its chemical structure, DIP is a relatively simple, planar
hydrocarbon without heteroatoms. In contrast to the intensely studied perylene
derivative 3,4,9,10-perylene tetracarboxylic dianhydride (PTCDA, $\mathrm{C}_{32}\mathrm{H}_{8}\mathrm{O}_6$)~\cite{hauschild_prl05,gerlach_prb07,henze_ss07,duhm_oe08,ziroff_prl10,Fenter_1997_PhysRevB}
with its four keto groups, the DIP--substrate interaction is not complicated by
polar side groups, and the influence of intermolecular interactions is
expected to be smaller than for PTCDA~\cite{kilian_prl08}.
%
Here, we present a systematic study with high-precision experimental data and
state-of-the-art calculations of DIP adsorbed on Cu(111), Ag(111), and
Au(111). This allows us to assess the role and relative contribution of the
vdW interactions, which, contrary to simplistic pictures, we find here to be
lowest for the most polarizable substrate.

\section{Computational approach}
\label{sec:comp_approach}
%
%
DFT calculations were performed using a method that extends standard pairwise
vdW approaches~\cite{tkatchenko_prl09,grimme_jcc06} to model adsorbates on
surfaces~\cite{ruiz_prl12}.  This was achieved by combining the DFT+vdW
scheme~\cite{tkatchenko_prl09} with the Lifshitz-Zaremba-Kohn (LZK) theory for
vdW interaction between an atom and the surface~\cite{Lifshitz,zaremba_prb76}
of a solid.
In our approach (DFT+vdW$^{\rm{surf}}$), the vdW energy is given by a sum of
$C_{6}^{ab} R^{-6}_{ab}$ terms, where $R_{ab}$ are the distances between atoms
$a$ and $b$, in analogy to standard pairwise dispersion corrected DFT methods. However, by
employing the LZK theory we include the many-body collective response
(\textit{screening}) of the substrate electrons in the determination of the
$C_6$ coefficients and vdW radii, going effectively beyond the pairwise
description. Interface polarization effects are accounted for via the
inclusion of semi-local hybridization due to the dependence of the $C_6^{ab}$
interatomic coefficients on the electron density in the DFT+vdW method. The
DFT+vdW$^{\rm{surf}}$ method has been shown to yield remarkably accurate
results for the structure and adsorption energies of xenon, benzene, and PTCDA
on a variety of (111) metal surfaces~\cite{ruiz_prl12,liu_prb12}. 
The FHI-aims code~\cite{blum_cpc09} was employed for our DFT calculations. The
repeated-slab method was used to model all systems with the vacuum gap set to
20 {\AA}.  In all calculations, convergence criteria of 10$^{-5}$ electrons
for the electron density and 10$^{-6}$ eV for the total energy of the system
were used. A convergence criterion of 0.01 eV/{\AA} for the maximum final
force was used for all structure relaxations. The scaled zeroth-order regular
approximation (ZORA) was applied for inclusion of scalar relativistic
effects~\cite{vanLenthe_1994_JChemPhys}. The DFT+vdW$^{\mathrm{surf}}$ method
employed the Perdew-Burke-Ernzerhof (PBE) functional~\cite{perdew_prl96}. The sampling of the Brillouin
zone was done using a $(2\times2\times1)$ $k$-point grid.

We used a $(7\times7)$ unit cell composed of a metal surface of three layers
and one single DIP molecule. In
the absence of experimental data for the in-plane registry we placed the
central ring of the molecule aligned with a topmost metal layer atom and the
major axis of the molecule aligned along the cell diagonal.  This structure
was adopted for the Cu(111), Ag(111), and Au(111) surfaces.  In each simulation
we obtained the adsorption energy curve using a rigid DIP molecule and tuning
the surface--molecule distance $d$.  The adsorption energy per molecule
$E_\mathrm{ads}$ was calculated from
$E_\mathrm{ads}=E_\mathrm{tot}-(E_\mathrm{surf}+E_\mathrm{DIP})$, where
$E_\mathrm{surf}$ is the energy per unit cell of the isolated metal surface,
$E_\mathrm{DIP}$ is the energy per unit cell of the isolated DIP molecule, and
$E_\mathrm{tot}$ is the energy per unit cell of the combined system.  We also obtained the relaxed geometries for all three systems starting from the
static equilibrium geometry.  During geometry relaxation, we allowed only the
topmost metal layer and the molecule to relax while the other two metal layers
were fixed.  From the final relaxed configurations we obtained the bonding
distance $d$ by taking the average position of all DIP carbon atoms with
respect to the unrelaxed topmost surface layer. This definition is consistent
with the analysis of the XSW data.

\section{Experimental details}
\label{sec:exp_details}
To measure the bonding distance of DIP we used the XSW
technique~\cite{woodruff_rpp05}, which yields precise and element-specific
structural data.  The experiments were performed at beamline ID32 of the ESRF
\cite{zegenhagen_jersp10}. DIP films were prepared and studied \textit{in
  situ} under ultrahigh-vacuum conditions. A separate preparation
chamber 
contained a Knudsen cell, a quartz crystal microbalance, installations for
Ar$^{+}$ sputtering, and a temperature-controlled sample stage. The main
chamber,
in which the XSW measurements were performed, was equipped with a sample
manipulator and a hemispherical SPECS PHOIBOS 225 HV photoelectron
analyzer. The XSW experiments were carried out at room temperature in
back-reflection geometry using the (111) Bragg reflection of the crystals for
at least two films per substrate to check for reproducibility of the results.
The detection angle of the analyzer was $\sim$90$^\circ$ relative to the
surface normal with an acceptance angle of $\pm 7.5^\circ$. We note that in
this configuration non-dipolar contributions to the photoelecton yield can be
effectively avoided~\cite{schreiber_ssl01}.
The Cu(111), Ag(111), and Au(111) single crystals were mounted on different
sample holders for individual treatment. The surfaces were prepared by
repeated cycles of Ar$^{+}$ bombardment and annealing at 700\,K. Surface
cleanliness was confirmed with x-ray photoelectron spectroscopy (XPS) as well
as low-energy electron diffraction (LEED). Sublimation grade DIP was
evaporated from a home-built Knudsen cell.
The intensity ratio of the C 1\textit{s} signal relative to a substrate core-level,
normalized with the corresponding photoemission cross sections, was used to
determine the number of DIP molecules on the surface. With the unit cell size
of DIP on Cu(111)~\cite{Oteyza_pccp09}, Ag(111)~\cite{Huang_pccp11}, and
Au(111)~\cite{Oteyza_jpcc08}, the coverages were calculated to be between
$0.3$ and $0.9$ ML.

\begin{figure}[t]
  \centering
  \includegraphics[width=.9\columnwidth]{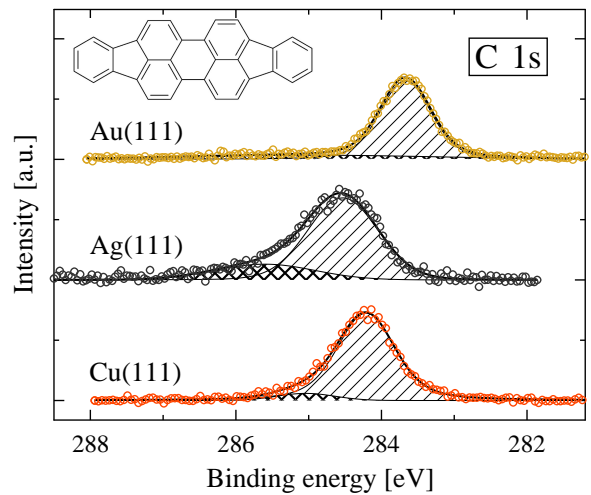}
  \caption{(Color online) C\,1s core-level shift observed for a submonolayer of DIP (inset)
    on Cu(111), Ag(111), and Au(111). The spectra were taken at an emission
    angle of 45$^{\circ}$ with the XSW setup at ID32. From each signal a
    Shirley background was subtracted and then fitted with a Voigt function
    for the main peak and a Gaussian function for possible shake-up peaks.}
  \label{fig:xps_fit}
\end{figure}

\section{Results and Analysis}
\label{sec:results}
\subsection{Experimental results}
\label{sec:results-exp}
The C 1\textit{s} core-level signals of DIP on Cu(111), Ag(111), and Au(111), which
were used for the XSW measurements, are shown in Fig.~\ref{fig:xps_fit}. The
main peaks are expected to consist of two principal components (C--C vs\ C--H
bound atoms) which, however, could not be resolved with the energy resolution of
the XSW setup. 
\begin{figure}[tbp]
  \centering
  \includegraphics[width=.9\columnwidth]{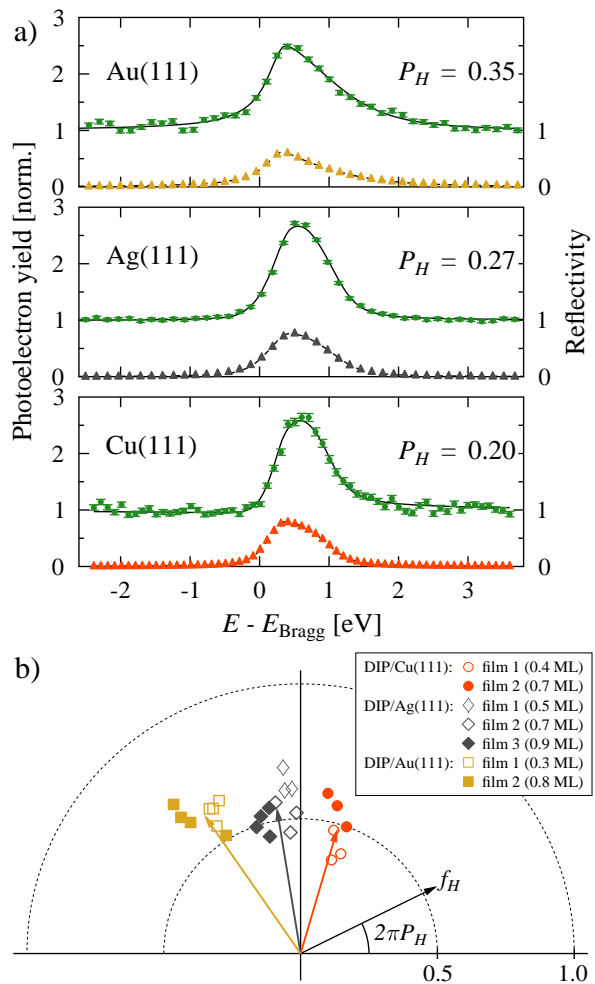}
  \caption{(Color online) (a) Typical XSW data for DIP showing the reflectivity (triangles)
    and photoelectron yield (circles) of the C 1\textit{s} signal on Cu(111), Ag(111),
    and Au(111). The solid lines correspond to least-squares fits of the reflectivity and photoelectron yield, which reveal the coherent position $P_H$ and coherent fraction $f_H$.  Bragg
    energies are $E_\mathrm{Bragg} = 2.97$ keV [Cu(111)] and
    $E_\mathrm{Bragg} = 2.63$ keV [Ag(111) and Au(111)]. (b) All XSW results
    for DIP on Cu(111), Ag(111), and Au(111) displayed in an Argand
    diagram. Here, each datapoint represents one single XSW measurement
    yielding $f_H$ (length of a vector) and $P_H$
    (angle of a vector). The three vectors point to the average values of $f_H$
    and $P_H$ for DIP on Cu(111), Ag(111), and Au(111). Film~1 of DIP on
    Ag(111) was measured with a different crystal compared to films 2 and 3.}
  \label{fig:dip_yield}
\end{figure}
In addition to each main peak, a second weak feature at
$\sim$1\,eV higher binding energy possibly related to a shake-up process can be
observed. Obviously, the binding energy of the C 1\textit{s} main line of DIP follows
$E_\mathrm{B}^{\mathrm{Ag}}>E_\mathrm{B}^{\mathrm{Cu}}>E_\mathrm{B}^{\mathrm{Au}}$,
being 284.5 eV on Ag(111), 284.2 eV on Cu(111), and 283.7 eV on Au(111).
Furthermore, the C 1\textit{s} peak of DIP on Ag(111) exhibits a stronger asymmetry
than on Cu(111) and Au(111).\footnote{This may be related to the creation of
  electron-hole-pairs close to the Fermi level caused by the higher density of
  states near the Fermi edge of DIP adsorbed on Ag(111) (Refs.~\onlinecite{Krause_diss09} and \onlinecite{Krause_2013_OrgElectron}).}
A detailed discussion of the spectroscopic features is beyond the scope of this paper in which we focus on the XSW results.
%
%

Representative results of the XSW experiments are shown in
Fig.~\ref{fig:dip_yield}(a). In each panel the measured reflectivity of the
substrate and the corresponding C 1\textit{s} photoelectron yield is
displayed. Least-squares fits of the data give the coherent position $P_H$ and
hence the average bonding distance $d_H=d_0(1+P_H)$~\cite{zegenhagen_ssr93},
where $d_0$ is the substrate lattice plane spacing. Based on results of all
XSW experiments we calculate the average bonding distance $d_H$ and the
standard deviation; see Fig.~\ref{fig:dip_yield}(b). For Cu(111) we thus find $(2.51
\pm 0.03)$ \AA{}, and for Ag(111) $(3.01 \pm 0.04)$ \AA{}. Due to the
reconstruction of the Au(111) surface, which results in a 3\% larger spacing
between the first and second Au layers~\cite{henze_ss07}, the bonding distance
decreases from the measured apparent value $(3.17 \pm 0.03)$ \AA{}\ to $(3.10
\pm 0.03)$ \AA. All experimental results are summarized in
Table~\ref{tab:xsw_results}.
%
Note that although the coverage of the two (three) DIP films prepared on each
substrate was not identical, we did not observe a significantly coverage-dependent bonding distance $d_H$.
\begin{table}[t]
  \centering
  \caption{Results of XSW experiments: Coherent fraction $f_H$, coherent
    position $P_H$, and bonding distance $d_H$ of DIP on the three noble
    metals. The parameters refer to an average of several XSW measurements
    with the corresponding standard deviation as error bars.}
  \begin{ruledtabular}
    \begin{tabular}{l|c|c|c}
      & $f_H$& $P_H$ &$d_H$  \\
      \hline
      Cu(111) & $0.48 \pm 0.09$ & $0.20 \pm 0.01$ & $(2.51 \pm 0.03)$\,\AA{} \\
      Ag(111) & $0.55 \pm 0.08$ & $0.28 \pm 0.02$ & $(3.01 \pm 0.04)$\,\AA{}  \\
      Au(111) & $0.62 \pm 0.06$ & $0.35 \pm 0.01$ & $(3.17 \pm
      0.03)$\,\AA{}\footnote{By taking the surface reconstruction of Au(111) into
        account, $d_H$ is reduced to 3.10 \AA{}.}\\
    \end{tabular}
  \end{ruledtabular}
  \label{tab:xsw_results}
\end{table}

Comparing these results with the bonding distances of PTCDA on the same metal
surfaces, i.e., $d_H=2.66$ \AA{} on Cu(111),\cite{gerlach_prb07}, $d_H=2.86$ \AA{} on Ag(111),\cite{hauschild_prl05, gerlach_prb07}, and $d_H=3.27$ \AA{} on Au(111),\cite{henze_ss07} we see that the bonding distances follow
the same order, i.e.,\ $d_H$(Cu) $ < d_H$(Ag) $ < d_H$(Au). Moreover, the
results demonstrate that the absence of the C=O groups in DIP affects the
bonding distance of the molecule only weakly.

\subsection{Computational results}
\label{sec:results-theo}
\begin{table}[htbp]
  \caption{Adsorption energy $E_\mathrm{ads}$ of the relaxed structures,
    vdW$^\mathrm{surf}$ binding energy $E_\mathrm{vdW}$ in parentheses as
    derived from data shown in Fig.~\ref{fig:adsorption}, and distances $d$
    between    the topmost layer of the metal and the carbon backbone of
    DIP. $d_\mathrm{min/max}$ refer to the lowest/highest bonding distances of
    a carbon atom within a DIP molecule.}
  \begin{ruledtabular}
    \begin {tabular}{l|c|c|c|c}
      & $E_\mathrm{ads}$ ($E_\mathrm{vdW}$) & $d$ & $d_{\mathrm{min}}$ & $d_{\mathrm{max}}$ \\
      \hline
      Cu(111) & -4.74 (-5.28)\,eV & 2.59\,\AA{} & 2.38\,\AA{} & 2.79\,\AA{} \\
      Ag(111) & -3.55 (-4.56)\,eV & 2.94\,\AA{} & 2.89\,\AA{} & 3.01\,\AA{} \\
      Au(111) & -2.53 (-3.06)\,eV & 3.22\,\AA{} & 3.15\,\AA{} & 3.29\,\AA{} \\
    \end{tabular}
  \end{ruledtabular}
  \label{tab:dft_results}
\end{table}
Having established precise experimental data, we now turn to the results of
our DFT calculations.  The average bonding distances of DIP obtained from
fully relaxed structures are $d=2.59$\,\AA \ on Cu(111), $d=2.94$\,\AA \ on
Ag(111), and $d=3.22$\,\AA \ on Au(111), see Table~\ref{tab:dft_results} and
Fig.~\ref{fig:adsorption}. 
We hence find that the PBE+vdW$^{\textrm{surf}}$
method applied to DIP on Cu(111), Ag(111), and Au(111) yields an agreement
better than 0.1\,\AA{} between theoretical calculations and experiments. In
accordance with the bonding distances, the calculated adsorption energies
listed in Table~\ref{tab:dft_results} follow the trend
$|E_\mathrm{ads}\mathrm{(Cu)}| > |E_\mathrm{ads}\mathrm{(Ag)}| >
|E_\mathrm{ads}\mathrm{(Au)}|$. Interestingly, Fig.~\ref{fig:adsorption} shows
that on Cu(111) the Pauli repulsion sets in rather weakly [a less steep
$E_\mathrm{ads}(d)$ for small distances] compared to Ag(111) and Au(111), which is due to significant interaction between DIP and Cu(111). One may speculate that the interaction mechanism includes hybridization between DIP and Cu states.

In addition to the adsorption energies and average bonding distances,
Table~\ref{tab:dft_results} holds the minimal and maximal bonding distances
$d_{\mathrm{min/max}}$ of individual carbon atoms in DIP. These values
indicate that the molecule adsorbs in a slightly tilted or distorted geometry.
For Cu(111), where the effect is most pronounced, the calculated bonding
distances $d_{\mathrm{min}}$ and $d_{\mathrm{max}}$ differ by
$\sim$0.4\,\AA{}, which is equivalent to a molecular tilt angle of
1.5$^\circ$.
The corresponding spread of vertical positions of the carbon atoms leads to a
reduced $f_H$ in the XSW scans. Model simulations similar to those presented
in Ref.~\onlinecite{Gerlach_2011_PhysRevLett} show that the DFT-derived
adsorption geometry on Cu(111) results in a relatively small decrease of the
coherent fraction ($\Delta f_H= -0.07$), which lies within the
standard deviation of our XSW measurements.

To obtain a better understanding of the influence of lateral intermolecular
interactions on the DIP adsorption geometry, we also computed the relaxed DIP
geometry for different Cu(111), Ag(111), and Au(111) unit cells. For DIP on
Cu(111), we increased the unit cell from $(7\times7)$ to $(9\times7)$ in order
to reduce the molecule--molecule interactions. We studied various
configurations, finding a flat relaxed geometry for each case considered. The
bonding distance is slightly larger (2.64 \AA) than for the calculation with
the smaller unit cell. For DIP on Ag(111), we also considered a unit cell
which was determined from a closed packed monolayer on
Ag(111)~\cite{Huang_pccp11}. The relaxed geometry of the molecules in the
monolayer is flat and the bonding distance $d=2.99$ \AA \ in almost
perfect agreement with the experimental one, i.e.,\ even better than the result
for Ag(111) shown in Table~\ref{tab:dft_results}. For a $(9\times5)$ unit cell
of Au(111), the relaxed DIP geometry yields an equilibrium distance of
3.15 \AA{}, also in slightly better agreement with experiment than the result
shown in Table~\ref{tab:dft_results}. Overall, these calculations agree with
the experimental observation that the vertical DIP position depends only
weakly on surface coverage.
%

\section{Discussion}
\label{sec:discussion}
\begin{figure}[htbp]
  \centering
  \includegraphics[width=.95\columnwidth]{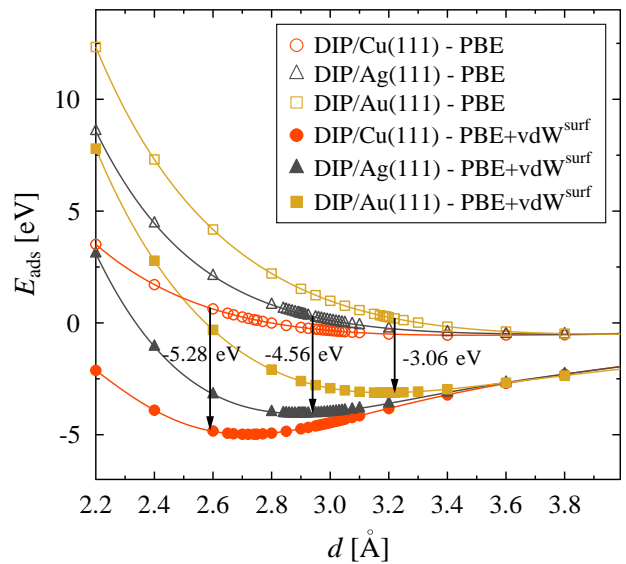}
  \caption{(Color online) Adsorption energy $E_{\mathrm{ads}}$
    for the unrelaxed DIP molecule as a function of its averaged distance $d$
    from the Cu(111), Ag(111), and Au(111) surfaces. The curves are shown for
    the PBE functional with and without the inclusion of long-range vdW
    interactions using the vdW$^{\rm{surf}}$ method.  The reported
    contribution of the vdW energy is shown at the equilibrium distance
    corresponding to the fully relaxed DIP--surface geometry (see text).}
  \label{fig:adsorption}
\end{figure}%

\begin{figure}[htb]
  \centering
  \includegraphics[width=.95\columnwidth]{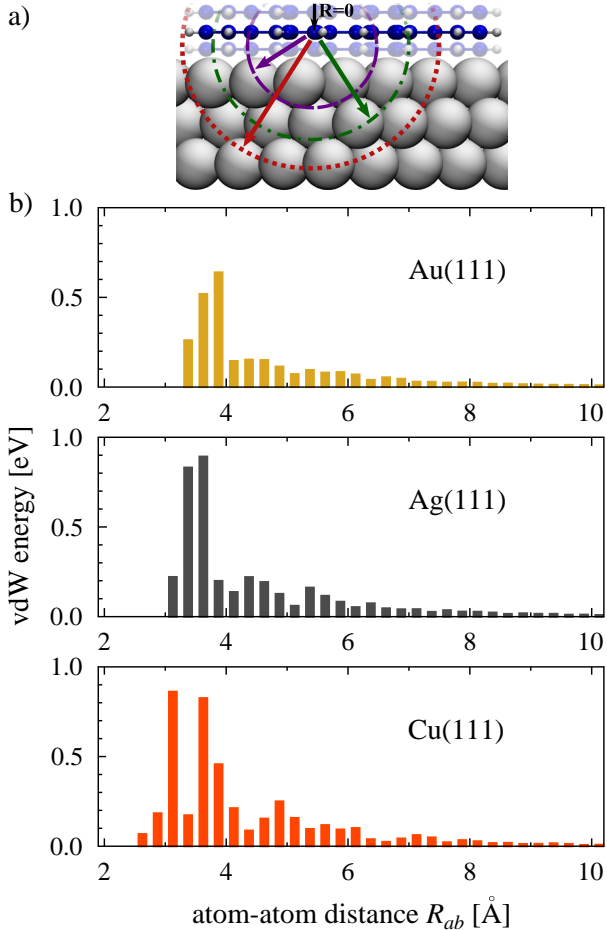}
  \caption{(Color online) (a) Schematic picture of the adsorption of DIP on a
      (111) crystal. Each arrow corresponds to a certain distance $R_{ab}$
      between adsorbate and substrate atoms. (b) vdW energy of DIP adsorbed on
      Cu(111), Ag(111), and Au(111) as a function of the distance between
      adsorbate and substrate atoms.}
  \label{fig:E_vdw}
\end{figure}

%
%
With the experimental and theoretical values at hand, and in view of their
excellent agreement, we are in a good position to discuss the vdW interactions
and the bonding distances in more detail. As described above, the (atom-atom)
vdW energy is computed as $C_6^{ab} R_{ab}^{-6}$, where the $C_6^{ab}$
coefficient determines the strength of the interaction between atoms $a$ and
$b$, while $R_{ab}$ is the distance between adsorbate and
  substrate atoms (Fig.~\ref{fig:E_vdw}). Integration of the vdW energy for a single atom adsorbed on a semi-infinite surface yields
the atom--surface vdW energy as~\cite{Bruch_2007_RevModPhys,Bruch_2009_Book} $C_3^{\textrm{A-S}} (z-z_0)^{-3}$, where now
$C_3^{\textrm{A-S}}$ determines the interaction strength between atom and
surface, $z$ corresponds to the distance of the atom to the uppermost surface
layer, and $z_0$ indicates the position of the surface image plane. In a
rather naive picture, the $C_3^{\textrm{A-S}}$ coefficient can be determined
simply from the $C_6^{aa}$ and the $C_6^{bb}$ coefficients that correspond to
the adsorbed atom and the metal atom, respectively. However, the situation
for real surfaces is more complex because both localized and bulk metal
electrons contribute to the $C_3^{\textrm{A-S}}$ coefficient in a non-trivial
way, meaning that this 
coefficient depends on the dielectric function of the underlying solid. We
computed the $C_3^{\textrm{A-S}}$ coefficients corresponding to the
interaction between a carbon atom and the Cu(111), Ag(111), and Au(111)
surfaces. When describing the metal surface as a simple collection of
non-interacting atoms we obtain $C_3^{\textrm{C-Cu}}=0.68$,
$C_3^{\textrm{C-Ag}}=0.55$, and $C_3^{\textrm{C-Au}}=0.50$
hartree$\cdot$bohr$^3$. In contrast, when using the more appropriate
Lifshitz-Zaremba-Kohn expression~\cite{Lifshitz,zaremba_prb76} for
$C_3^{\textrm{A-S}}$, we obtain $C_3^{\textrm{C-Cu}}=0.35$,
$C_3^{\textrm{C-Ag}}=0.35$, and $C_3^{\textrm{C-Au}}=0.33$
hartree$\cdot$bohr$^3$. This clearly illustrates that the vdW interaction
between an atom and a solid surface is significantly modified by the
collective electronic response within the substrate surface~\cite{ruiz_prl12,Bruch_2007_RevModPhys,Bruch_2009_Book}.
The very similar LZK $C_3$ coefficients for Cu, Ag, and Au lead to essentially
the same adsorption energy at large distances for DIP on Cu(111), Ag(111), and
Au(111) (Fig.~\ref{fig:adsorption}). However, at shorter
molecule--surface distances, which include the equilibrium distance, the
adsorption energy is determined by an interplay between the vdW attraction and
the Pauli repulsion with a possible covalent component. The Pauli repulsion
follows roughly the trend of decreasing vdW radii, with a faster onset in
terms of the molecule--surface distance for Au (with the largest vdW radius), and
then decreasing for Ag and Cu. Therefore, for Au the balance between the vdW
attraction and the Pauli repulsion is obtained {\it further away} from the
substrate (i.e.,\ at larger adsorption distances) than for Cu, which in turn makes
the adsorption energies {\it lower} for Au than for Cu, in contrast to the
possible naive expectation of Au with its higher polarizability and $C_6$
coefficient exhibiting a stronger vdW interaction than Cu.

The difference in the vdW energy distribution for DIP on Cu(111), Ag(111), and Au(111) is visualized in Fig.~\ref{fig:E_vdw}(b), where the vdW energy between DIP and substrate atoms is plotted as a function of their distance $R_{ab}$. In contrast to Ag(111) and Au(111), the small bonding distance of DIP on Cu(111) results in a second peak in the histogram at $\sim$3.6 \AA{}, which originates from the higher atomic density
of the Cu substrate. 

\section{Conclusion}
\label{sec:conclusion}

In conclusion, the bonding distances calculated with the PBE+vdW$^{\rm{surf}}$
method are in excellent agreement with the XSW data for DIP on Cu(111),
Ag(111), and Au(111) (2.51, 3.01, and 3.10 \AA{},
respectively). Our combined study demonstrates that the vdW energy is larger
for DIP on Cu(111) than for DIP on Ag(111) and Au(111). Future investigations on the electronic properties of these systems, which can draw on the findings presented here, will contribute to an even better understanding of the adsorption process. 

\section*{Acknowledgments}

This work was financially supported by the DFG (SCHR700/14-1 and SFB951) and MEXT. N.F.\ and
A.T.\ are grateful for support from the FP7 Marie Curie Actions of the EU, via
the Initial Training Network SMALL (Grant No. MCITN-238804). The authors gratefully acknowledge the ESRF for providing access to beamline ID32.
%

\bibliographystyle{aipnum4-1}
%
%

\end{document}